%% file: main.tex
\documentclass[preprint,journal]{vgtc}         

\ifpdf
  \pdfoutput=1\relax                   
  \usepackage{graphicx}                
  \DeclareGraphicsExtensions{.pdf,.png,.jpg,.jpeg} 
\else
  \usepackage{graphicx}                
  \DeclareGraphicsExtensions{.eps}     
\fi%

\graphicspath{{figures/}{pictures/}{images/}{./}} 

\usepackage{microtype}                 
\PassOptionsToPackage{warn}{textcomp}  
\usepackage{textcomp}                  
\usepackage{mathptmx}                  
\usepackage{times}                     
\usepackage{cite}                      
\usepackage{tabu}                      
\usepackage{booktabs} 
\usepackage{adjustbox}
\usepackage{amsmath}

\usepackage{subfig}
\usepackage{diagbox}
\usepackage{boldline}
\usepackage{url}
\usepackage[bookmarks=false]{hyperref}
\usepackage{tikz}
\usepackage{multirow}
\usepackage{array}
\usepackage{boldline}
\usepackage{xcolor} 
\usepackage{float}
\usepackage{subfig}
\usepackage{bm}
\usepackage{tikz}
\usepackage{longtable}
\usepackage{color, colortbl}
\usepackage{float}
\usepackage{amssymb}
\usepackage{appendix}
\usepackage{cancel}

\onlineid{1068}

\vgtccategory{Research}

\vgtcpapertype{Technology}

\title{A privacy-preserving approach to streaming eye-tracking data}

\author{Brendan David-John, \textit{Student Member, IEEE}, Diane Hosfelt, \\ Kevin Butler \textit{Senior Member, IEEE} and Eakta Jain, \textit{Member, IEEE}}
\authorfooter{
\item
 Brendan David-John is a PhD student at the University of Florida. \\ E-mail: brendanjohn@ufl.edu.
\item
 Diane Hosfelt was a privacy and security researcher at Mozilla at the time of writing E-mail: dianehosfelt@gmail.com
\item
 Dr. Kevin Butler is an Associate Professor at the University of Florida. \\ E-mail: butler@ufl.edu
\item
 Dr. Eakta Jain is an Assistant Professor at the University of Florida. \\ E-mail: ejain@cise.ufl.edu
}

\shortauthortitle{John-David \MakeLowercase{\textit{et al.}}: short title here}

\abstract{
Eye-tracking technology is being increasingly integrated into mixed reality devices. Although critical applications are being enabled, there are significant possibilities for violating user privacy expectations. We show that there is an appreciable risk of unique user identification even under natural viewing conditions in virtual reality. This identification would allow an app to connect a user's personal ID with their work ID without needing their consent, for example. To mitigate such risks we propose a framework that incorporates gatekeeping via the design of the application programming interface and via software-implemented privacy mechanisms. Our results indicate that these mechanisms can reduce the rate of identification from as much as 85\% to as low as 30\%. The impact of introducing these mechanisms is less than 1.5$^\circ$ error in gaze position for gaze prediction. Gaze data streams can thus be made private while still allowing for gaze prediction, for example, during foveated rendering. Our approach is the first to support privacy-by-design in the flow of eye-tracking data within mixed reality use cases.

}


\CCScatlist{ 
 \CCScat{K.6.1}{Management of Computing and Information Systems}%
{Project and People Management}{Life Cycle};
 \CCScat{K.7.m}{The Computing Profession}{Miscellaneous}{Ethics}
}

\teaser{
   \centering
    \includegraphics[width=\linewidth]{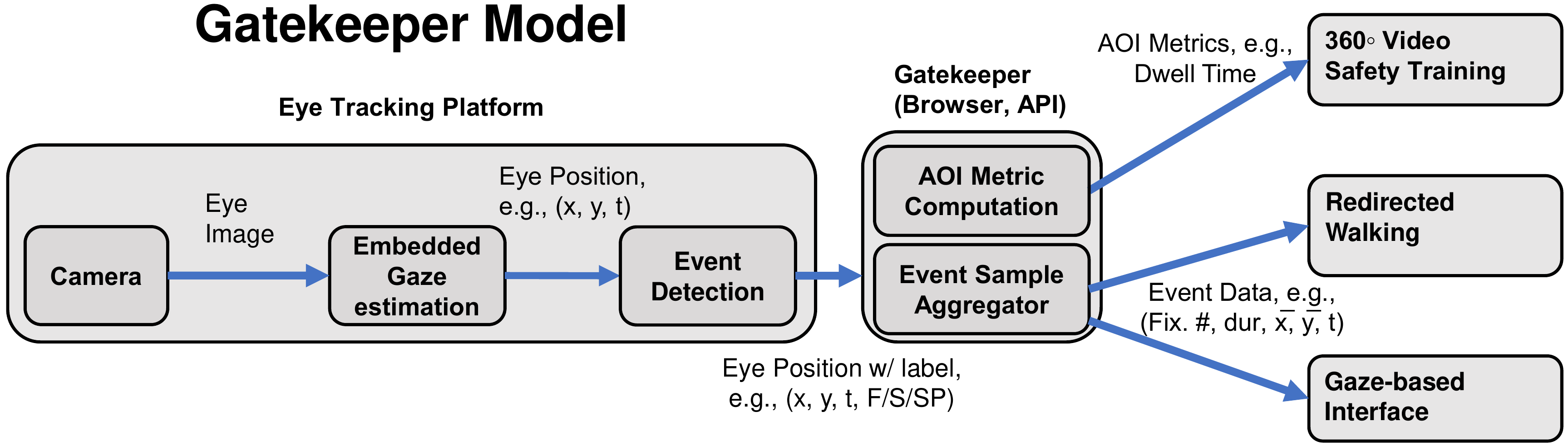}
    \includegraphics[width=\linewidth]{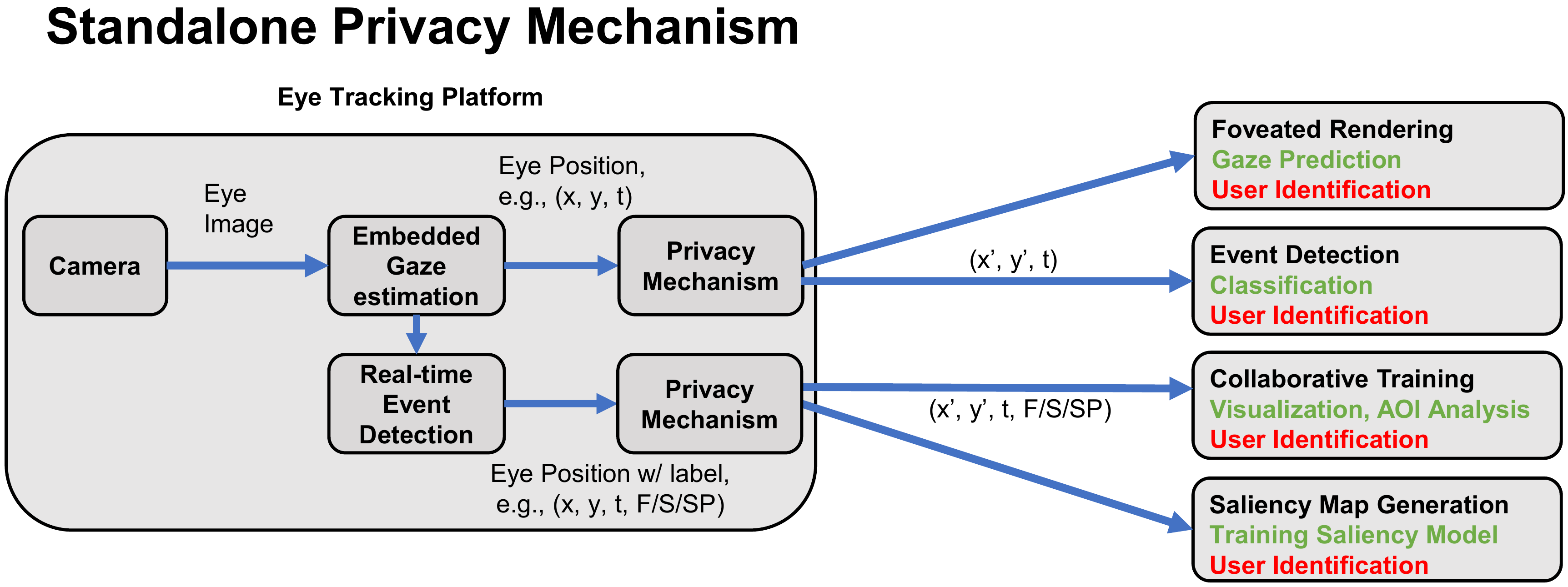}
    \caption{Top: The \textit{Gatekeeper} model protects identity by delivering relevant data at different levels directly through the API, while withholding raw gaze samples that contain biometric features. This approach cannot be used directly with applications that require raw gaze samples. Bottom: In scenarios where a \textit{Gatekeeper} API cannot be implemented, we instead apply a privacy mechanism to raw gaze samples to serve applications that use gaze samples or event data directly. 
    }
    \label{fig:dataflow-overview}
}

\nocopyrightspace

\vgtcinsertpkg

\newcommand{\mytilde}{\raise.17ex\hbox{$\scriptstyle\mathtt{\sim}$}}

\usepackage{algorithm}
\usepackage[noend]{algpseudocode}
\newfloat{algorithm}{t}{lop}

\renewcommand\copyrighttext{%
  \footnotesize \textcopyright 2021 IEEE. Personal use of this material is permitted.
  Permission from IEEE must be obtained for all other uses, in any current or future
  media, including reprinting/republishing this material for advertising or promotional
  purposes, creating new collective works, for resale or redistribution to servers or
  lists, or reuse of any copyrighted component of this work in other works.
  DOI: \href{}{10.1109/TVCG.2021.3067787}}
\newcommand\copyrightnotice{%
\begin{tikzpicture}[remember picture,overlay]
\node[anchor=south,yshift=10pt] at (current page.south) {\fbox{\parbox{\dimexpr\textwidth-\fboxsep-\fboxrule\relax}{\copyrighttext}}};
\end{tikzpicture}%
}

\begin{document}

\newcolumntype{?}{!{\vrule width 1.5pt}}
\newcolumntype{P}[1]{>{\centering\arraybackslash}p{#1}} 

\maketitle

\copyrightnotice

\input{introduction.tex}
\input{api.tex}
\input{methodology.tex}

\input{results.tex}
\input{conclusion.tex}

\acknowledgments{Authors acknowledge funding from the National Science Foundation\,(Awards FWHTF-2026540, CNS-1815883, and CNS-1562485), the National Science Foundation GRFP\,(Awards DGE-1315138 and DGE-1842473), and the Air Force Office of Scientific Research\,(Award FA9550-19-1-0169).}

\bibliographystyle{abbrv-doi}

\bibliography{refs,vr2019}



\end{document}

%% file: introduction.tex
\section{Introduction}

As eye trackers are integrated into mixed reality hardware, data gathered from a user's eyes flows from the mixed reality platform to the applications\,(apps) that use this data. This data is a critical enabler for a number of mixed reality use cases: streaming optimization~\cite{lungaro2018gaze}, foveated rendering~\cite{patney2016towards,bastani2017foveated,meng2018kernel,meng2020eye}, redirected walking~\cite{sun2018towards,langbehn2018blink,keyvanara2019transsaccadic,keyvanara2020effect}, gaze-based interfaces~\cite{rajanna2018gaze,zhang2019accessible,gebhardt2019learning}, education~\cite{rahman2020exploring}, and social interaction~\cite{lombardi2018deep,mousas2019effects,macquarrie2019perception,cho2020effects,murcia2020evaluating}. The eye-tracking data also contains a variety of information about the user which are not necessarily needed by each application. For example, eye movements identify attributes such as gender, bio-markers for various health conditions, and identity. As a result, how this data is handled, and to whom, has privacy and security implications.

The problem of applications receiving data and passing it along to colluding apps or parent companies erodes public trust in technology, and cannot be ``regulated away''. It has received public attention in the context of similar personal devices, such as smartphones. Recently, The Weather Channel took location data it mined from users’ foot traffic at different businesses, and sold it to hedge funds to inform their investments before quarterly income statements were released.\footnote{\url{https://www.nytimes.com/interactive/2019/12/19/opinion/location-tracking-cell-phone.html}}. Even with regulation, imagine that the weather app collecting location data colludes with an advertising application that belongs to the same parent company. The user will then be served personalized ads based on her location: such as car ads appearing after a visit to the car dealership for an oil change. Now imagine that the parent company also knows which cars she glanced at while waiting, or that she actually spent most of the time looking at the motorcycle parked out front relative to the other vehicles.

This problem becomes even more severe when we recognize that mixed reality headsets are going to have as much enterprise use as personal use. A user might log in at work to do their job-related training with their known real-world identity, but attend labor union meetings as User X to avoid negative repercussions.\footnote{\url{https://tcf.org/content/report/virtual-labor-organizing/}}$^,$\footnote{\url{https://www.foley.com/en/insights/publications/2015/09/be-careful-what-you-say-during-a-union-organizing}} The agent that connects these two identities has the power to ``out'' the user to her work organization.

In this paper, we have investigated the threat of biometric identification of a user from their eye movements when they are being eye tracked within immersive virtual reality environments. For several mixed reality use cases, raw eye-tracking data does not need to be passed along to the application. As shown in Figure 1, a \textit{Gatekeeper} that resides between the eye tracking platform and applications can alleviate this threat by encapsulating raw data within an application programming interface\,(API). We have proposed a design for such an API in Section\,\ref{sec:api}.

This philosophy of serving data on a ``need-to-know basis'' is effective in preventing data from being used for deviant purposes instead of their originally intended purpose. However, there remain certain applications that rely on access to raw gaze data. In this case, we have proposed privacy mechanisms to erase identifying signatures from the raw gaze data before it is passed on to the application. We have evaluated how the proposed privacy mechanisms impact utility, i.e., what the application needs gaze data to do. Finally, we have investigated how the proposed privacy mechanisms impact applications that need access to eye events, i.e., eye-tracking data labeled as fixations, saccades, or smooth pursuits.

Our work is part of a broader thrust in the eye tracking and virtual reality communities on characterizing risks related to unregulated massive scale user eye tracking, and developing technological mitigations for these risks. For risks associated with an adversary gaining access to the eye image itself, we direct readers to the privacy mechanisms presented in~\cite{John2019,chaudhary2020privacy}. For a differential privacy perspective, we direct readers to~\cite{steil2019privacy,liu2019differential,john2020let}. For a differential privacy perspective on the identification of users by colluding apps, we direct readers to the detailed analysis in~\cite{steil2019privacy,bozkir2020differential}, with the caveat that the utility task considered in this body of work is gaze-based document type classification. In contrast, we focus on utility tasks that are specific to mixed reality. Our goal is to provide a foundation for future researchers and developers to organize their thinking around the risks created by the flow of behavioral data in mixed reality, and the proactive rather than reactive design of mitigation strategies.

\section{Eye-Tracking Applications in Mixed Reality}
\label{sec:services}
We can expect eye tracking to run as a service within a mixed reality device, analogous to the way that location services run on phones today. Eye tracking is a specific case of more general behavioral tracking services in mixed reality, including head, hand, and body tracking. Mixed reality platforms such as Microsoft and Facebook will collect raw data from the native sensors, process it to perform noise removal and event detection, and pass the processed data up the software stack. Because a rich, self-sustaining mixed reality ecosystem will rely on independent content developers, a mixed reality web browser, akin to a conventional web browser, will provide the software interface to access a wide array of content for consumers. In this section, we highlight critical eye-tracking applications for mixed reality that use aggregate-level, individual-level, and sample-level gaze data. 

\subsection{Aggregate-level eye-tracking applications}
Aggregate gaze data is collected from many viewers to drive applications such as highlighting salient regions using heatmaps~\cite{rai2017dataset,david2018dataset,sitzmann2018saliency}, and learning perceptual-based streaming optimizations for 360$^\circ$ content~\cite{lungaro2018gaze,xu2020state}. These applications typically rely on a data collection process conducted in research lab environments for a sample of viewers. Viewer data is then used to train machine-learning models or evaluate the most effective streaming methodology within the dataset. Results from the dataset are then released in aggregate form to inform the deployment of such methods on consumer devices. This provides utility to the consumer without creating privacy risks, however training data for machine-learning models may pose a risk to privacy~\cite{chen2020machine}, as well as publicly-released datasets that include the raw gaze data used to generate aggregate representations~\cite{li2018bridge,xu2018gaze,hu2019sgaze,agtzidis2019ground,hu2020dgaze}.

\subsection{Event-level eye-tracking applications}
Eye movement behavior captured by eye-tracking events, such as fixations, saccades, and smooth pursuit, contribute to gaze-based interfaces~\cite{pai2017gazesphere,gebhardt2019learning}, evaluating training scenarios~\cite{duchowski2000binocular,burova2020utilizing,john2020look}, and identifying neurodegenerative diseases~\cite{orlosky2017emulation} and ASD~\cite{bradshaw2019use}. Detecting eye-tracking events enables improved techniques for redirected walking~\cite{langbehn2018blink,keyvanara2019transsaccadic,keyvanara2020effect}, a critical application for VR that expands the usable space of virtual environment within a confined physical environment. The most common method to quantify an individual's gaze behavior is to mark Areas of Interest\,(AOIs) within content and measure how gaze interacts with this region. Typical metrics for these regions depend on fixation and saccade events only, recording dwell times, the number of fixations or glances, and fixation order~\cite{le2013methods,orquin2016areas}. Event data also poses a privacy risk, as it reveal the viewer's intent and preferences based on how gaze interactions with different stimuli content.

\begin{table*}[t]
    \centering
    \caption{State-of-the-art gaze-based biometric methods. Key: RBF = Radial Basis Function Network, RDF = Random Decision Forests, STAT = Statistical test, SVM = Support Vector Machine.}
     \label{tab:methods}
    \begin{tabular}{|l|c|c|c|c|}
    \hline 
     \textbf{Method} &\textbf{Features}& \textbf{Classifier}& \textbf{Dataset} & \textbf{Results}   \\ \hline 
     Schroder et al.~\cite{schroder2020robustness} &Fixation, Saccade& RBF& BioEye 2015, MIT data set & IR: 94.1\%, 86.76\%  \\ \hline
     Schroder et al.~\cite{schroder2020robustness} &Fixation, Saccade& RDF& BioEye 2015, MIT data set & IR: 90.9\%, 94.67\%  \\ \hline
     George\&Routray\cite{george2016score}&Fixation, Saccade & RBF& BioEye 2015& IR:93.5\% \\ \hline
     Lohr et al.~\cite{lohr2020eye} &Fixation, Saccade& STAT& VREM-R1, SBA-ST & EER: 9.98\%, 2.04\%   \\ \hline
     Lohr et al.~\cite{lohr2020eye} &Fixation, Saccade& RBF& VREM-R1, SBA-ST & EER: 14.37\%, 5.12\%   \\ \hline 
     Eberz et al.~\cite{eberz201928}&Fixations, Binocular Pupil & SVM& \cite{eberz201928}& EER: 1.88\% \\ \hline
     Rigas et al.~\cite{rigas2016towards}& Fixations, Saccades, Density maps& Multi-score fusion & \cite{rigas2016towards}&EER: 5.8\%, IR: 88.6\% \\ \hline 
     Monaco~\cite{monaco2014classification}& Gaze Velocity/Acceleration& STAT&EMVIC 2014&IR: 39.6\% \\ \hline


    \end{tabular}
       
    \end{table*}

\subsection{Sample-level eye-tracking applications}
Multiple key mixed reality applications depend on individual gaze samples from an eye-tracker of a sampling rate of at least 60Hz. This includes foveated rendering~\cite{patney2016towards,bastani2017foveated,meng2018kernel,meng2020eye}, which is expected to have the biggest impact on deploying immersive VR experiences on low-power and mobile devices. This application relies on gaze samples to determine where the foveal region of the user currently is, and to predict where it will land during an eye movement to ensure that the user does not perceive rendering artifacts~\cite{arabadzhiyska2017saccade}. Similarly, gaze prediction models are trained that predict future gaze points while viewing 360$^\circ$ imagery and 3D rendered content~\cite{hu2019sgaze,hu2020dgaze}.

Another key set of applications that require sample-level data are gaze guidance techniques~\cite{rothe2018gazerecall,rothe2019guidance}. Gaze guidance takes advantage of sensitivity to motion in the periphery to present a flicker in luminance that will attract the user's eyes, using eye tracking to remove the flicker before the user can fixate upon the region and perceive the cue~\cite{bailey2009subtle,grogorick2017subtle}. This technique enables manipulation of visual attention, and ultimately user behavior. For example, gaze guidance in 2D environments has been shown to improve spatial information recall~\cite{bailey2012impact}, improve training of novices to identify abnormalities in mammogram images~\cite{sridharan2012subtle}, and improve retrieval task performance in real-world environments~\cite{booth2013guiding}. Gaze guidance has also been used to enhance redirected walking techniques in VR by evoking involuntary eye movements, and taking advantage of saccadic suppression~\cite{sun2018towards}. Guiding gaze through saccades and manipulating the user allows for use of a 6.4m$\times$6.4m virtual space within a 3.5m$\times$3.5m physical space, significantly improving upon the usable area within VR experiences. This application requires an eye tracker sampling rate of 250Hz or more, and requires sample-level data to know precisely when gaze moves towards the periphery cue. Providing sample-level data with high accuracy at this frequency poses a serious risk to user privacy in the form of gaze-based biometric features that can then be extracted from these gaze positions. 

\section{Related Work}
\label{ref:section-privacy-risks}


Human eyes reflect their physical attributes. For example, algorithms can estimate the ages of users by monitoring the change in the gaze patterns as they age~\cite{munoz1998age,zhang2018old}, their gender based on the temporal differences in gaze patterns while viewing faces~\cite{sammaknejad2017gender}, and their race from the racial classification of faces they tend to look at~\cite{bar2006nature}.

Beyond physical attributes, gaze allows rich insights into psychological attributes, such as neurological~\cite{leigh2015neurology} and behavioral disorders~\cite{dalton2005gaze,pelphrey2005neural,mundy2018review}. The eyes can also reveal whether an individual suffers from an affective disorder---anxious individuals' gaze is characterized by vigilance for threat during free viewing, while depressed individuals' gaze is characterized by reduced maintenance of gaze on positive stimuli~\cite{armstrong2012eye}. Eye tracking has also been used to investigate gaze behavior in individuals on the autism spectrum, finding that they generally tend to fixate less on faces and facial features~\cite{boraston2007application,chawarska2009looking}. 

Pupillometry, when combined with scene metadata could allow algorithms to infer user sexual orientation, as shown in clinical studies measuring genital responses, offering a less invasive way to infer individual's preferences~\cite{rieger2015sexual}. In addition to allowing sexual orientation inferences, pupillometry can reveal insight into women's hormonal cycles using similar methodology~\cite{laeng2007women}. Pupil size also reveals the user's cognitive load~\cite{duchowski2018index} as well as emotional arousal, as shown in studies with images~\cite{lang:1993, bradley:2009} and videos~\cite{raiturkar2016decoupling}. Interestingly, pupil response seems to be modulated by subconscious processing, changing when the mind wanders~\cite{uzzaman2011eyes}.

Body mass index\,(BMI) status appears to influence gaze parameters that are not under conscious control, allowing BMI estimation when presenting individuals with images of foods of differing caloric content~\cite{graham2011body}. These risks involve knowledge of both eye position and stimuli, whereas user identification can be applied to raw eye movements without knowledge of what the stimuli was.

\subsection{State-of-the-art in user identification based on eye movements}
Gaze patterns can be used to identify individuals as they contain unique signatures that are not under a user's voluntary control~\cite{kasprowski2012first, kasprowski2014second}. The Eye Movement Verification and Identification Competitions in 2012 and 2014 challenged researchers to develop algorithms that identified users based on their eye movements when they followed a jumping dot\,(2012) and when they looked at images of human faces\,(2014). The best models' accuracy ranged from 58\% to 98\% for the jumping dot stimuli, and nearly 40\% accuracy compared to a 3\% random guess probability for viewing faces.

Based on recent surveys on eye movements biometrics~\cite{galdi2016eye,rigas2017current} as well as our own literature search, we identified algorithms that have been shown to successfully identify individual users from their eye movements in Table\,\ref{tab:methods}. These algorithms have been applied to existing gaze-biometric challenge datasets, as well as the natural viewing of image stimuli in 2D\,(MIT data set). The method with the best biometric performance produces an Equal Error Rate of 1.88\% using pupil-based features~\cite{eberz201928}, however the majority of consumer applications in mixed-reality do not require pupil diameter. Thus, we selected to implement the RBF approach proposed by George and Routray~\cite{george2016score}, as it relies only on fixation and saccade events. This method also produces impressive results with VR eye-tracking data~\cite{lohr2020eye} and natural viewing of 2D images~\cite{schroder2020robustness}.

\subsection{State-of-the-art in eye-tracking security and privacy}
In recent years privacy concerns related to eye-tracking applications has grown significantly~\cite{kroger2019does,steil2019privaceye,John2019,john2020security,bozkir2020privacy,kaleido2020}. In response, researchers have developed methods to enhance privacy of aggregate features, like saliency heatmaps~\cite{liu2019differential} and event statistics~\cite{steil2019privacy,bozkir2020differential,fuhl2020reinforcement}. These methods have been shown to reduce performance in classification of gender and identity, however the methods operate only on aggregate gaze data after it has been collected and processed. Recent work by Li et al. has applied formal privacy guarantees to raw streams of gaze designed to obfuscate viewer's gaze relative to AOIs within stimuli over time~\cite{kaleido2020}. The ability to protect biometric identity was was evaluated empirically on the 360\_em dataset~\cite{agtzidis2019ground}, reducing identification to chance rate. 
Our work develops a threat model based on the streaming of gaze samples and the privacy risk related to biometric identification within an XR ecosystem.

%% file: api.tex
\section{Designing an API for Gaze Privacy}
\label{sec:api}

The typical architecture and data flow in an eye-tracking platform is shown in Figure~\ref{fig:dataflow-overview}. Existing eye trackers process user data in three stages: eye image capture, which images the user's eye, eye position estimation, which infers the point of regard from the eye image, and event detection, which classifies each point of regard as belonging to a fixation, saccade, blink, etc. When eye trackers were specialty equipment, all this data was made available to the application. These applications were typically research data gathering software. The major difference now is that the applications will have a profit-based business model. This model will naturally create incentives to share user gaze data and make inferences by combining data across devices for advertising revenue, for example. We have identified privacy risks created by this ecosystem in Section~\ref{ref:section-privacy-risks}. In this section, we define our threat model and propose the design of an application programming interface\,(API) which adopts a privacy-preserving approach to passing gaze data to downstream applications.

 \noindent \textbf{Threat Model}
We assume that the components comprising the eye-tracking platform and API are trusted, i.e., the integrity of the hardware and software could be attested through mechanisms such as secure boot~\cite{arbaugh1997secure} and integrity measurement~\cite{sailer2004design}, and we assume that the operating system is protected, e.g., through SELinux mandatory access controls~\cite{morris2002linux}. The adversary is capable of examining all data transmitted to the eye-tracking applications, and seeks to use this information to re-identify the user. An adversarial application has the capability to collude with other applications by sharing information through either overt or covert channels~\cite{marforio2012analysis} in order to re-identify users. 

Our privacy-preserving solution is focused on preventing biometric identification of users from their gaze data. First, the eye is imaged by a camera, producing an eye image that is provided to the platform, which processes the image into position coordinates. The platform provides this eye position to trusted applications like the browser, which then pass the eye position on to browser apps that perform tasks such as AOI analysis for performance in training scenarios, saccade detection for redirected walking, and smooth pursuits for gaze-based interaction.

 \noindent\textbf{Nai\"ve API Design}
The simplest way to provide a gaze API would be to pass along the raw gaze data to applications. At any point in time, the application would be able to request \texttt{getGazePosition()}. From this, the application would be able to compute fixations, saccades, and dwell time; in particular, an AOI application would be able to compute fixations in an AOI, time to first saccade into the AOI, and dwell time in the AOI.

Providing raw gaze data also allows for computation of the velocity of eye movements, and other features that are commonly used for identity classification tasks~\cite{george2016score,galdi2016eye,schroder2020robustness}. Allowing for raw gaze access in an untrusted context, such as the web, allows arbitrary apps the ability to re-identify users.

\subsection{Enabling AOI Metrics}
However, we can modify the gaze API to be privacy-preserving by acting as a \textit{Gatekeeper}. Privacy vulnerabilities are caused by the design assumption that the application is benign, and the data is used only for the purpose for which it is collected. 
As discussed previously, applications need not be benign, and connecting user data across devices will allow for richer inferences to be made about that user. This threat motivates our proposed \textit{Gatekeeper} design. An added benefit of our proposed design is that the \textit{Gatekeeper} model provides desired metrics directly to applications, instead of requiring applications to process streamed user gaze data and calculate the metrics themselves. 

Advertisers and other AOI applications are interested in the number of fixations and the dwell time of a fixation in a predetermined AOI. Under the \textit{Gatekeeper} framework, instead of passing along raw gaze positions, an API allows requests for this information. For example, a \texttt{getFixations} method takes a rectangular area and returns a list of fixations that had occurred in that area, and a \texttt{getDwellTime} method takes as input a fixation and returns in milliseconds the dwell time of the fixation. Additionally, we provide a \texttt{getSaccades} method that would return a list of saccades into the AOI. Saccades are a strong classifier feature for identity, when raw gaze points are included, however we mitigate this risk by providing only lower dimensional summary data. 

It is important to note that this API is designed specifically to provide AOI metrics and summary data of eye movement events. The API does not scale to address applications such as platform foveated rendering, which requires raw gaze samples for utility. The \textit{Gatekeeper} model does support streaming optimizations based on the current gaze position within a discrete set of tiles~\cite{chakareski2018viewport,papaioannou2019tile}, by providing only information about which tile they are currently attending too. This type of optimization is critical for low-power devices to ensure high visual quality while preserving precious network resources.

\subsection{Enabling Real-time Event Data}
In some situations, such as gaze-based interfaces and redirected walking, applications will need to be notified when a new fixation or saccade occurs, instead of querying for all fixations or saccades. 

In this scenario, we can use an \texttt{EventListener} model instead of a query-based model. When a new event occurs, the \texttt{EventListener} will be notified and given the event data, \texttt{(x, y, t)} and a boolean indicating if it is a fixation, saccade, or smooth pursuit. More complex eye movements are difficult to detect in real-time with the sampling rate of mixed reality eye-tracking devices, and typically are not implemented in real-time applications.

Our typical model for streaming event data is to send an event when the eye movement has concluded. For example, in a gaze-based interface the application needs to be notified that a smooth pursuit occurred, and where it landed. In applications such as redirected walking it is critical to know when a saccade begins, to take advantage of saccadic blindness~\cite{sun2018towards, langbehn2018blink, keyvanara2019transsaccadic, keyvanara2020effect}. In this case, one mode of the \texttt{EventListener} will be to indicate when a saccade event has started and finished, as opposed to only when the saccade has finished. 

\subsection{Enabling Privacy-enhanced Sample Data}
Most applications will be able to function with the aforementioned API designs; however, two key mixed reality applications that will require sample-level data are foveated rendering and subtle gaze guidance. 

Foveated rendering is critical for performance on next generation wearable VR headsets. In an ideal situation, platforms will use GPU-based foveated rendering---where gaze information is sent to the graphics driver, informing it to do fewer calculations for the parts of the screen that are away from the center of view. This requires cooperation with the graphics hardware driver for optimal performance. Experiments on native platforms show up to a 2.71 times speed up in frames per second~\cite{meng2020eye}. This will not be possible in all cases, so platforms and browsers will also need to leverage software-based foveated rendering and streaming optimization\cite{mueller2018shading}. In this scenario, gaze samples are transmitted directly to the content or webpage, which then knows where it should render objects in more detail. However, this exposes the raw gaze data to the application and allows the content to perform further processing on the raw gaze information, whether that is user identification or inferring sensitive characteristics.

In these scenarios the eye-tracking platform must stream sample-level data, and it is impossible to simply abstract data using a privacy-preserving API. Therefore, we propose the use of a privacy mechanism to manipulate gaze samples as they are streamed to increase privacy.

%% file: methodology.tex
\section{Methodology}
\label{sec:mechanisms}
In this section, we propose, implement, and evaluate three privacy mechanisms with the goal of mitigating the threats identified in Section~\ref{sec:api}. Our goal is to reduce the accuracy of user identification based on features derived from common eye events, such as fixations and saccades. We consider the following privacy mechanisms: addition of Gaussian noise to raw gaze data, temporal downsampling, and spatial downsampling.

We implement these mechanisms and evaluate them against the baseline identification rate when raw gaze data is passed to the application as is. For each of the privacy mechanisms, we also evaluate the utility of the data that is passed downstream.

\subsection{Privacy Mechanism Definitions}
\label{sec:privacy_mechanisms}

We define the data received by the privacy mechanism to be a time series where each tuple is comprised of horizontal and vertical gaze positions\,($x,y$), a time stamp $t$, and the event label assigned to the sample $e$: $X=\{(x_1,y_1,t_1,e_1), (x_2,y_2,t_2,e_2),..., (x_G,y_G,t_G,e_G)\}$, a set of $G$ gaze positions. This data is processed via a privacy mechanism and the processed output as a time series $X'$, with additional variables defined in Table\,\ref{tab:definitions}. The following three privacy mechanisms are explored in this paper.

\begin{table}[t]
    \centering
    \caption{Privacy mechanism variable definitions.}
    \label{tab:definitions}
    \begin{tabular}{|c|l|} \hline 
         Variable & Description \\ \hline
         $x$ & Horizontal gaze position \\ \hline 
         $y$ & Vertical gaze position \\ \hline
         $t$ & Timestamp\\ \hline
         $e$ & Event label: Fix.\,(F), Sacc.\,(S), Smooth Pursuit\,(SP)\\ \hline
         $X$ & Input time series of gaze samples\\ \hline 
         $G$ & Number of gaze positions in time series\\ \hline
         $X'$& Output privacy-enhanced time series\\ \hline
         $K$ & Temporal downsample factor relative to sampling rate\\ \hline
         $L$ & Spatial downsample factor relative to 3840$\times$2160\\ \hline
         $M$ & Number of rows in equirectangular projection\\ \hline
         $N$ & Number of columns in equirectangular projection\\ \hline
         $\delta_x$ & Horizontal step size: $\frac{360}{N}$ \\ \hline
         $\delta_y$ & Vertical step size: $\frac{180}{M}$ \\ \hline
    \end{tabular}
\end{table}
\begin{table*}[t]
    \centering
    \caption{Dataset characteristics.}
    \label{tab:datasets}
    \begin{tabular}{|c|c|c|c|c|c|c|} \hline 
         Dataset & Participants & \# Stimuli & Avg. \# Stimuli & Stimuli Duration &Stimuli Type& Task \\ \hline
         ET-DK2\,(ours) & 18 & 50 & 50 & 25s &360$^\circ$ Images& Free Viewing \\ \hline
         VR-Saliency~\cite{sitzmann2018saliency} & 130 & 23 & 8 & 30s& 360$^\circ$ Images & Free Viewing  \\ \hline
         VR-EyeTracking~\cite{xu2018gaze} & 43 & 208 & 148 & 20s-70s& 360$^\circ$ Videos & Free Viewing \\ \hline
         360\_em~\cite{agtzidis2019ground} & 13 & 14 & 14 & 38s-85s &360$^\circ$ Videos & Free Viewing \\ \hline
         DGaze~\cite{hu2020dgaze} & 43 & 5 & 2 & 180s-350s &3D Rendered Scene &Free Viewing \\\hline

    \end{tabular}
    
\end{table*}

\textbf{Additive Gaussian Noise} \,\, Noise is sampled from a Gaussian distribution of zero mean and standard deviation $\sigma$ defined in visual degree is added to the gaze positions. Noise is independently sampled for horizontal and vertical gaze positions as $X'=\{(x_1+N(0,\sigma),y_1+N(0,\sigma),t_1,e_1), (x_2+N(0,\sigma),y_2+N(0,\sigma),t_2,e_2),..., (x_G+N(0,\sigma),y_G+N(0,\sigma),t_G,e_G)\}$.

\textbf{Temporal Downsampling} \,\, Temporal downsampling reduces the temporal resolution of the eye-tracking data stream. Downsampling is implemented by streaming the data at a frequency of the original sampling rate divided by a scaling parameter $K$. 
The output time series is defined as $X'=\{(x_{(K\cdot p)+1},y_{(K\cdot p)+1},t_{(K\cdot p)+1},e_{(K\cdot p)+1}),...\}$ for all integers $p \in [0,\frac{G}{K}]$. 
For example, with a scaling parameter of two, the private gaze positions are defined as $X'=\{(x_1,y_1,t_1,e_1), (x_3,y_3,t_3,e_3), (x_5,y_5,t_5,e_5),...\}$, retaining only every other gaze sample. For a scaling parameter of three, $X'=\{(x_1,y_1,t_1,e_1), (x_4,y_4,t_4,e_4), (x_7,y_7,t_7,e_7),...\}$.

\textbf{Spatial Downsampling}
Spatial downsampling reduces the resolution of eye-tracking data down to a discrete set of horizontal and vertical gaze positions. Intuitively, the scene is divided into a grid and each gaze sample is approximated by the grid cell that it lies within. Spatial downsampling is performed by defining a target equirectangular domain spanning 180$^\circ$ vertically and 360$^\circ$ horizontally with $M$ rows and $N$ columns. For smaller values of $M$ and $N$ there are less possible positions, and thus reduced spatial resolution. Raw gaze positions $(x\in [0,360^\circ),y\in [0,180^\circ),t)$ are transformed by first computing the horizontal step size $\delta_y = \frac{180}{M}$ and vertical step size $\delta_x = \frac{360}{N}$. Downsampled gaze positions are then computed as $(\lfloor \frac{x}{\delta_x}\rfloor \cdot \delta_x, \lfloor \frac{y}{\delta_y}\rfloor \cdot \delta_y, t)$, where $\lfloor \cdot \rfloor$  represents the floor function that rounds down to the nearest integer.

For the results presented in this paper, we parameterize spatial downsampling as a factor $L$ relative to an equirectangular domain of $M=2160$ and $N=3840$, mapping to a domain of $M=\frac{2160}{L}$ and $N=\frac{3840}{L}$. For example, an input downsampling factor of $L$ equals two will result in $M=1080$ and $N=1920$, a factor of $L$ equals three will result in a resolution of $M=720$ and $N=1280$, and so on.

\subsection{Datasets}
In order to evaluate the privacy mechanisms on how effectively they prevented an adversary from re-identifying the user, we selected five existing datasets of VR eye-tracking data. Table\,\ref{tab:datasets} presents characteristics of each dataset included in analysis. Datasets were selected to have diversity in the number of participants, the number of stimuli presented, and the task being performed. Four of the datasets are publicly available, while ET-DK2 consists of data previously collected by the authors.\footnote{The dataset will be released publicly when the manuscript is published} 

\subsubsection{ET-DK2}
The ET-DK2 dataset consists of twenty participants viewing fifty 360$^\circ$ images using an Oculus-DK2 HMD with integrated SMI 60Hz binocular eye tracker. Data was collected under an IRB approved protocol in December 2017 for the purpose of generating saliency maps from gaze data. Two participants were not included in analysis, as one participant got motion sickness, and the data collection software did not log data from all 50 images for one participant. The remaining 18 individuals were made up of five females and thirteen males with an average age of 32, and an age range of 23 to 52 years. Each participant viewed images from the Salient360!\,\cite{rai2017dataset} dataset in random order. Participants were seated in a swivel chair so they could rotate and explore each 360$^\circ$ scene while eye and head movements were recorded.

All participants performed a 9-point calibration at the beginning of the experiment, and eye-tracking accuracy was validated to less than $2^\circ$ visual angle before image viewing. Each 360$^\circ$ image was shown for 25 seconds, following the Salient360!\,\cite{rai2017dataset} protocol. In contrast to their protocol, we varied the starting orientation of the participant within the 360$^\circ$ image across eight orientations instead of being held constant. Halfway through the experiment participants were given a five minute break, after which the eye tracker was re-calibrated before viewing the rest of the images. The entire data collection process took approximately 40 minutes, including informed consent and a post-study demographics survey.

\subsubsection{VR-Saliency}
The VR-Saliency~\cite{sitzmann2018saliency} dataset includes gaze data collected from participants viewing 360$^\circ$ images on a 2D display, in VR while seated in a swivel chair, and in VR while standing. We analyze only the seated VR condition, as it is the only VR condition with raw data available at 120Hz for all stimuli. Free-viewing data was collected in a similar manner to ET-DK2 for the purpose of saliency map generation, however only eight 360$^\circ$ images were viewed by each participant. 

\subsubsection{VR-EyeTracking}
The VR-EyeTracking~\cite{xu2018gaze} dataset includes gaze data collected at 100Hz from participants viewing 360$^\circ$ videos. The dataset application is to train a deep network model for predicting gaze within dynamic VR environments. The video stimuli did not have a fixed duration, as in ET-DK2 and VR-Saliency, however participants viewed many videos and took many breaks to avoid motion sickness.

\subsubsection{360\_em}
The 360\_em~\cite{agtzidis2019ground} dataset includes gaze data collected at 120Hz from participants viewing 360$^\circ$ videos. Fourteen of the stimuli consisted of typical 360$^\circ$ videos from YouTube, while one stimuli was created by the authors to elicit specific eye and head movements. The dataset application is to train and evaluate event detection algorithms, classifying fixation, saccade, smooth pursuit, and OKN events in VR viewing data. For our analysis we only consider the fourteen stimuli downloaded from YouTube.

\subsubsection{DGaze}
The DGaze~\cite{hu2020dgaze} dataset includes gaze data collected at 100Hz from participants that explore and navigate various 3D rendered scenes. Within each environment multiple animals dynamically move around, attracting visual attention of the participant. Gaze data is used to train and evaluate the DGaze model for gaze prediction. DGaze can predict gaze position given head orientation, or predict the next gaze position given the current gaze position. Gaze prediction by DGaze has been demonstrated in the context of foveated rendering, and can help account for latency in the eye-tracking and rendering pipeline~\cite{patney2016towards,arabadzhiyska2017saccade,hu2020dgaze}.

\subsection{Metrics}
For each dataset metrics are computed to identify privacy risks, and evaluate the impact of privacy mechanisms on application utility. Utility measures depend on the application of eye-tracking within the datasets, ranging from AOI analysis to gaze prediction. We define a utility metric for each dataset depending on the type of stimuli and application.

\subsubsection{Privacy}
In our context, privacy refers to how effectively the mechanism prevents an adversary from identifying an individual. Identification is defined as a classification task: an algorithm matches the input to the database and return the closest match.
If the algorithm matches the input to the ground truth identity, then the comparison is counted as a True Positive, otherwise it is considered a False Negative. The Identification Rate\,(IR), is the total number of True Positive classifications divided by the total number of comparisons~\cite{kasprowski2012first,kasprowski2014second,schroder2020robustness}. A high IR indicates accurate classification of identity, and therefore, low privacy.

\subsubsection{Utility}
Predicting future gaze position from eye-tracking data is a critical area of research that has yet to be solved~\cite{hu2019sgaze,hu2020dgaze}. Using the DGaze dataset we evaluate the ability to predict ground truth gaze position 100 ms into the future when gaze data output from a privacy mechanism is used as the testing data, and as both the training and testing data. Utility is measured as angular gaze prediction error for each input gaze sample, with lower values indicating higher accuracy. 

The most common form of eye-tracking analysis is performed using static AOIs defined within image content~\cite{le2013methods,orquin2016areas}. AOI analysis is used to study gaze behavior during social interaction~\cite{bennett2016looking}, while viewing websites~\cite{yangandul2018many}, and to evaluate content placement in 3D environments~\cite{alghofaili2019optimizing}, among many other applications. A key AOI metric that is robust to fixation detection parameters is dwell time\cite{orquin2016areas}. Dwell time measures how long a viewer's gaze fell within an AOI, and allows for comparison between which AOIs attracted the most attention. We evaluate the loss in utility between ground truth and gaze data output by a privacy mechanism by computing the Root Mean Squared Error\,(RMSE) between AOI dwell times. AOI utility is measured for the ET-DK2 dataset, as two rectangular AOIs are marked within each image that correspond with a salient object, such as people or natural landmarks, to measure individual viewing behavior within the scene.

Eye-tracking data is also used to generate saliency maps, which represent a probability distribution over visual content that highlights regions most likely to be looked at by a viewer~\cite{le2013methods}. Saliency maps generated from aggregate eye-tracking data from many viewers and are used to train and evaluate deep learning models for saliency and scanpath prediction~\cite{assens2018pathgan,chen2020salbinet360}. Saliency metrics are computed for both 360$^\circ$ images\,(VR-Saliency), and 360$^\circ$ video\,(VR-EyeTracking and 360\_em). We compute KL-Divergence~\cite{le2013methods} to measure the impact on aggregate-level gaze measures and saliency modeling.

\subsection{Implementation Details: Biometric Re-identification}
\label{sec:biometric_classification}
We define two classifiers for biometric identification using a Radial Basis Function\,(RBF) network~\cite{george2016score,lohr2020eye}, with one network to classify fixation events and one to classify saccade events. This method is analogous to a traditional neural network with an input layer representing a feature vector $\vec{x} \in \mathbb{R}^p$ containing $p$ fixation or saccade features from a single event, one hidden layer consisting of $m$ nodes, and an output layer containing $c$ class scores, one for each unique individual in the dataset. The output class scores are used to measure which individual the input feature vector is most similar to. Thus, larger scores indicate a higher probability of the fixation or saccade event being from that class, or individual. Each node in the hidden layer is defined by an activation function $\phi_i(\vec{x})$ and a set of real-valued activation weights $w_{i,c}$, where $i \in [1,2,\dots,m]$ and $j \in [1,2,\dots,C]$. The similarity score for a given class $c$ in the output layer is computed as a weighted sum of all activation functions in the hidden layer, 
\begin{equation}
    Score_c(\vec{x}) = \sum_{i=1}^{m} w_{i,c}\cdot \phi_i(\vec{x}).
\end{equation}

The activation function of each hidden node takes the form of a Gaussian distribution centered around a prototype vector $\vec{\mu}_i$ with spread coefficient $\beta_i$. The function is defined as 
\begin{equation}
    \phi_i(\vec{x}) = e^{-\beta_i ||\vec{x}-\vec{\mu}_i||^2},
\end{equation}


with shape coefficient $\beta_i$ and prototype feature vector $\vec{\mu}_i$ defined prior to training the network. Thus, an RBF network must be constructed in two stages by first defining the prototypes and then optimizing the activation weights.

First, k-means clustering is applied to a training set of $n$ feature vectors to determine $k$ representative feature vectors per individual~\cite{george2016score,lohr2020eye}. Through this process $\beta_i$ and $\vec{\mu}_i$ are defined for each of the $m = k\cdot c$ hidden nodes. The activation function $\phi_i(\vec{x})$ is then defined using the cluster centroid as $\vec{\mu}_i$, and $\beta_i$ as $\frac{1}{2\sigma}$, where $\sigma$ is the average distance between all points in the cluster and the centroid $\vec{\mu}_i$. 

Second, the activation weights $w_{i,c}$ are learned from the same set of training data used to define the activation functions. Weights are trained using only fixation or saccade features from the training set. Training can be implemented using gradient descent~\cite{schwenker2001three}, or by the Moore–Penrose inverse when setting up the network as a linear system~\cite{george2016score}. The latter method is implemented in this work by defining the RBF network using an activation output matrix $A_{n\times m}$, where rows consist of the $n$ training feature vectors input to the $m$ previously defined activation functions, weight matrix $W_{m \times c}$ comprised of activation weights $w_{i,c}$, and an output matrix $Y_{n\times c}$ generated as a one-hot encoding of the ground truth identity labels. Using matrix multiplication the following system defines the RBF Network $A \cdot W = Y$.

The weight matrix $W$ is then learned by computing $W = A^* \cdot Y$, where $A^*$ is the Moore-Penrose inverse of $A$ computed using MATLAB's $pinv$ implementation. Class score predictions $\hat{Y}$ are then generated for the testing data $\hat{A}$ by computing $\hat{A} \cdot W = \hat{Y}$. Every sample in the testing set is then classified as the class label with the maximum score. To classify a stream of events the class scores from all events are first summed together, and then the class with the maximum value returned. Scores from the fixation RBF and saccade RBF are combined by averaging the score between the two and summing them together for equal contribution to the final classification. 

\begin{figure}[t]
    \centering
    \includegraphics[width=\linewidth]{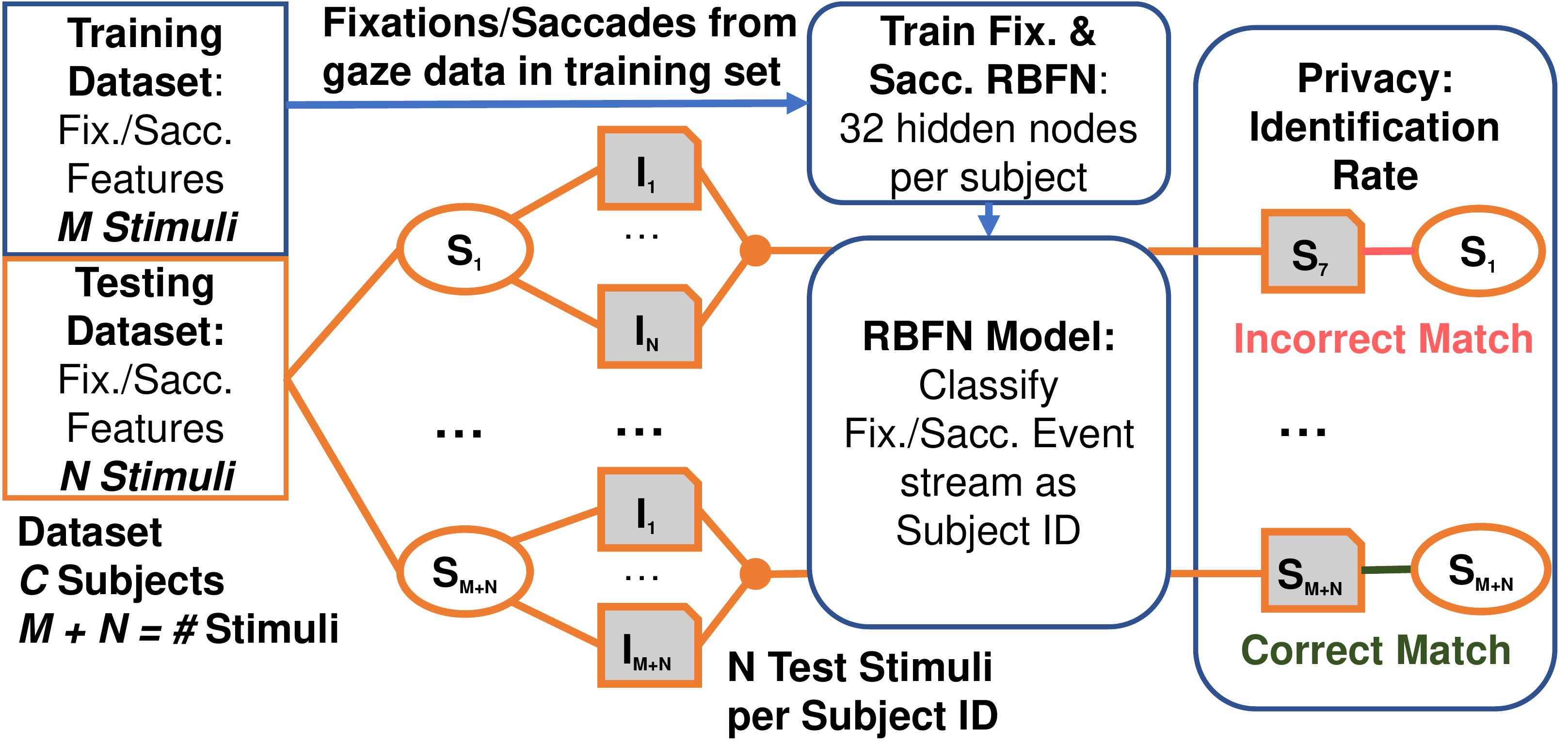}
    \caption{Evaluation procedure for the gaze-based biometric classifier.}
    \label{fig:evaluation}
\end{figure}

\begin{figure}[t]
    \centering
    \includegraphics[width=\linewidth]{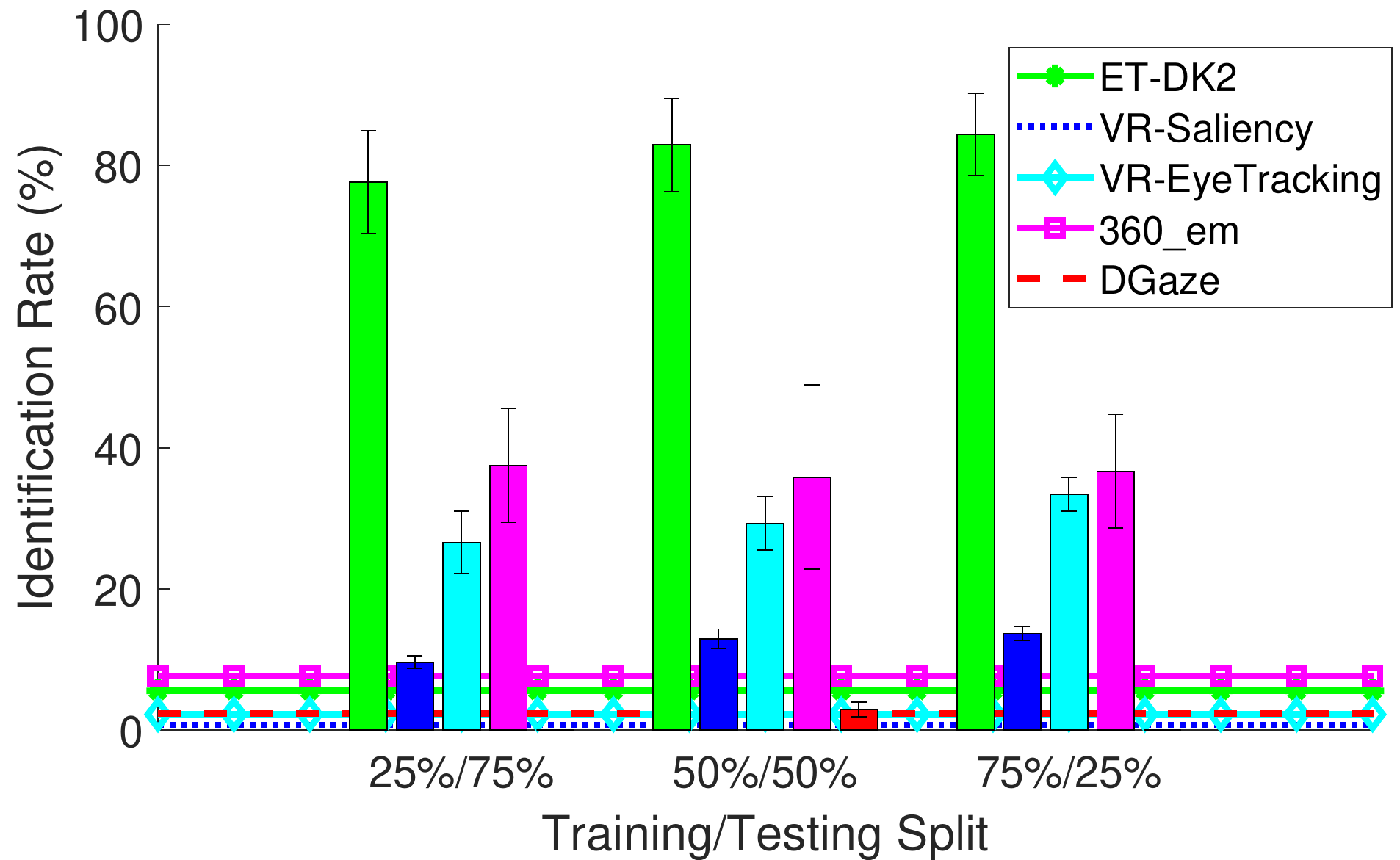}
    \caption{Mean and standard deviations of identification rates across datasets of 360$^\circ$ images\,(ET-DK2, VR-Saliency), 360$^\circ$ videos\,(VR-EyeTracking, 360\_em), and 3D rendered scenes\,(DGaze). Lines for each dataset indicate a baseline of random guessing for the given number of subjects.}
    \label{fig:all_results}
\end{figure}

\begin{figure*}[t]
    \centering
    
    \includegraphics[width=0.33\linewidth]{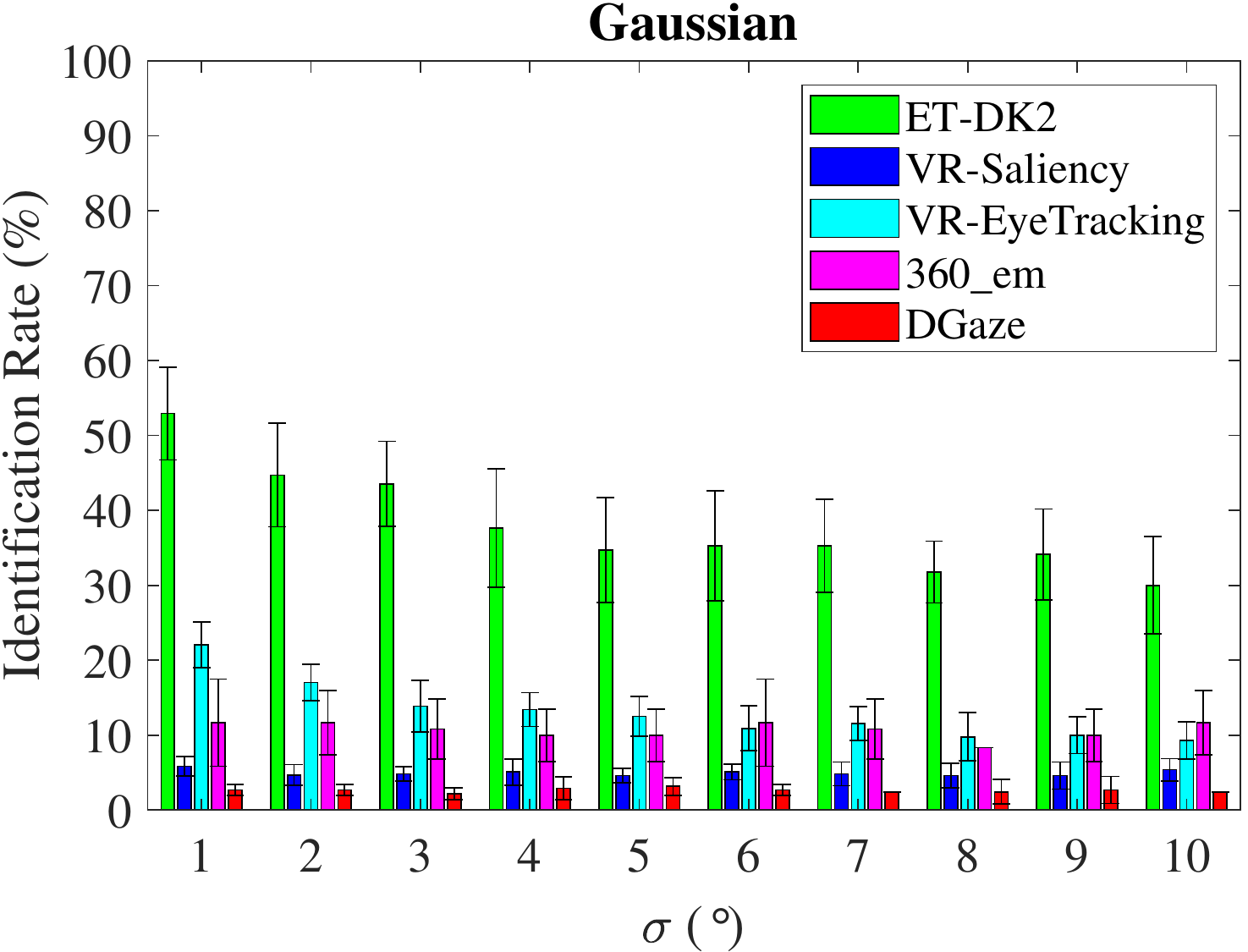}\includegraphics[width=0.33\linewidth]{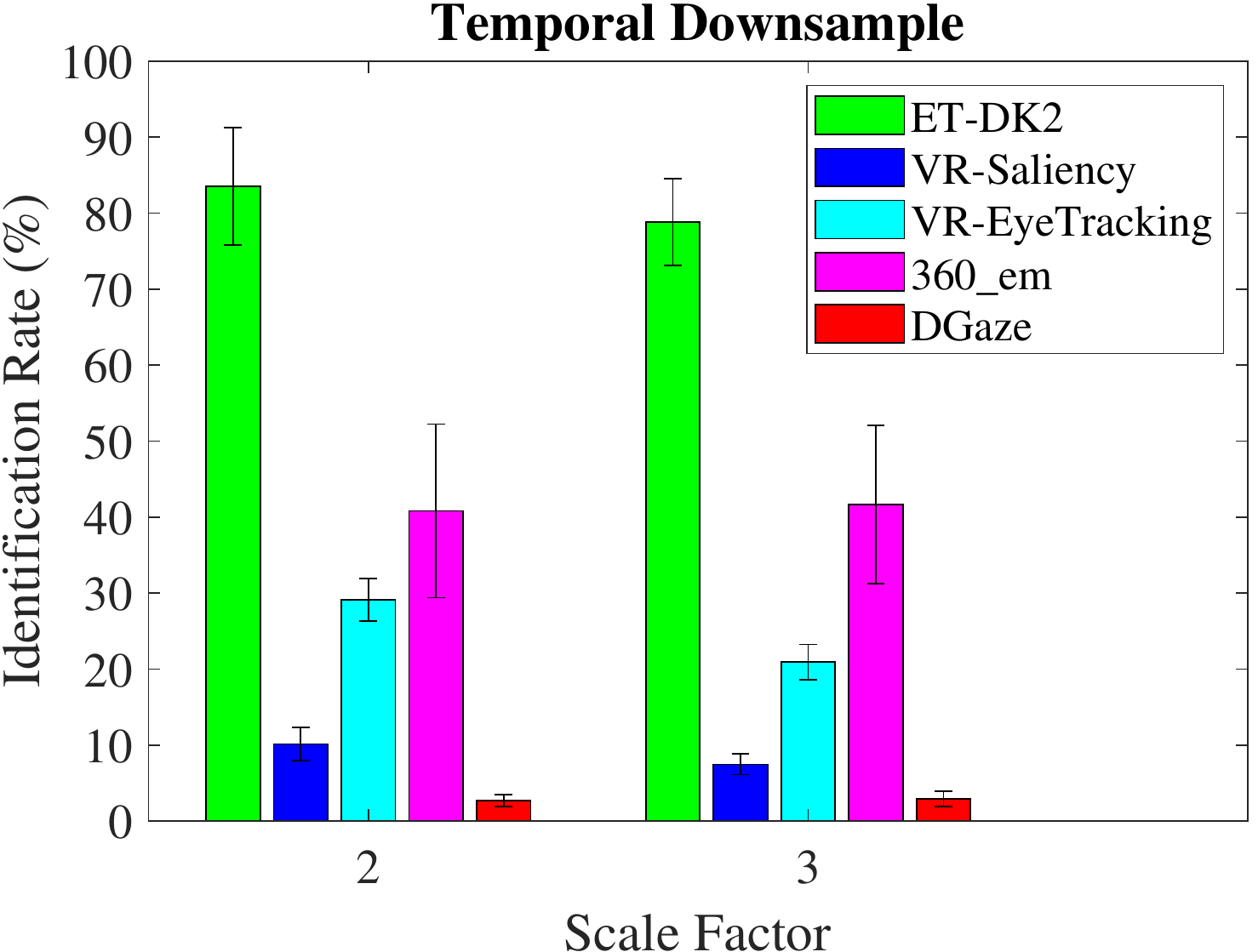}\includegraphics[width=0.33\linewidth]{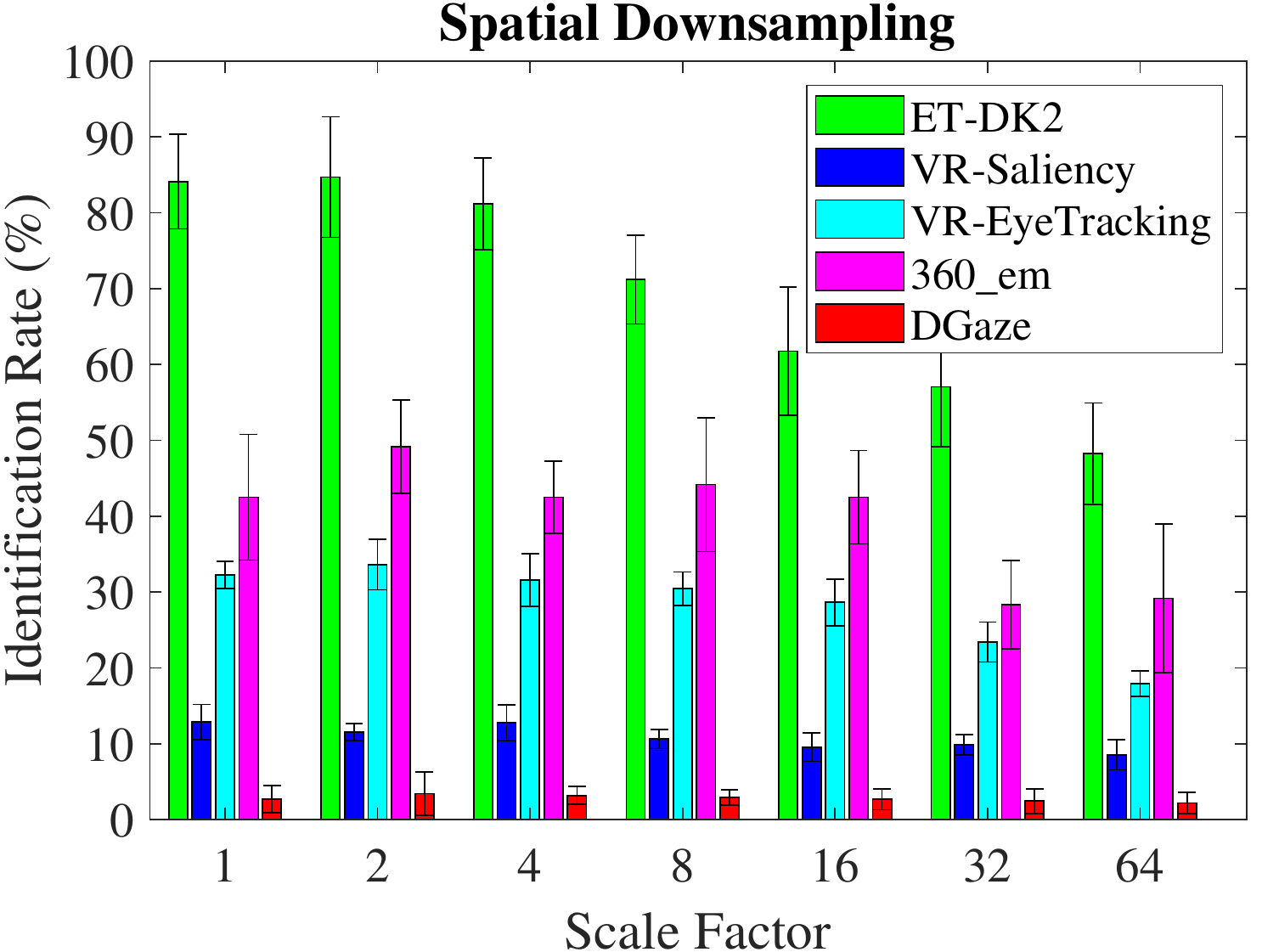}
    \caption{Mean and standard deviation of identification rate for each privacy mechanism with different internal parameters. Gaussian noise generates the lowest observed identification rates across all datasets, while temporal downsampling has the least impact.}
    \label{fig:privacy_mechanism_results}
\end{figure*}

\subsection{Evaluation Protocol}
\label{sec:evaluation}
The evaluation protocol for the RBF-based biometric, illustrated in Figure\,\ref{fig:evaluation}, is derived from~\cite{schroder2020robustness}, where a stream of gaze data collected from multiple participants viewing numerous static images is used for training and testing the identity classification. The size of the training and testing sets are defined by the number of stimuli from which gaze data is used. For example, with a training/testing split of 50\%/50\% gaze data from a half of the dataset selected at random is used for training and the other half for testing. Fixation and saccade events data from all $C$ participants are aggregated from the training stimuli and are then used to train the fixation and saccade RBF networks for classifying identity, as described in Section\,\ref{sec:biometric_classification}. Fixation and saccade events from the testing set are input to the trained RBF networks to classify the identity of each participant. Each participant is present in both the training set and the testing set. Identification rate is then computed as the number of correct matches divided by the number of comparisons.

%% file: results.tex
\section{Results}
In this section we will compute privacy and utility metrics to evaluate the proposed privacy mechanisms from Section\,\ref{sec:privacy_mechanisms} for each dataset listed in Table\,\ref{tab:datasets}. In Section\,\ref{sec:biometric_results}, we first compute identification rate using the RBF biometric for each dataset without modification, to establish a baseline privacy risk. Then, we compute identification rate for the privacy mechanisms for different parameter values and discuss observed effects. Last, in Section\,\ref{sec:utility_results} we explore the privacy achieved by each mechanism, and the measured impact on eye-tracking utility.

\begin{table*}[!h]
\caption{ This table illustrates the impact of introducing the Gaussian Noise privacy mechanism on the identification rate as well as on three use cases. The reported numbers are for $\sigma=10^\circ$. The second column shows how the identification rate falls after the privacy mechanism is applied. The fourth column reports an error metric that is relevant to that use case.} \label{tab:tradeoff_gaussian}
\begin{tabular}{|p{1.66cm}|c|c|c|c|} \hline
    Mechanism & Identif. Rate & Utility & Impact on Utility & Dataset  \\ \hline
    \begin{tabular}{@{}l@{}}Gaussian \\ Noise\end{tabular} & 3\% $\rightarrow$ 2\% & Gaze Prediction & Avg. Prediction Error Difference = $1.14^\circ$ & DGaze (Re-trained) \\ \hline
    \begin{tabular}{@{}l@{}}Gaussian \\ Noise\end{tabular}  & 85\% $\rightarrow$ 30\% & AOI Analysis & Dwell Time RMSE = 0.0359s & ET-DK2 (360$^\circ$ images) \\ \hline
    \begin{tabular}{@{}l@{}}Gaussian \\ Noise\end{tabular}  & 33\% $\rightarrow$ 9\% & Generate Saliency Map & KL-Divergence = 0.0367 & VR-EyeTracking (360$^\circ$ videos) \\ \hline
    
\end{tabular}
\end{table*}
\begin{table*}[!h]
\caption{This table illustrates the impact of introducing the Temporal Downsample privacy mechanism on the identification rate as well as on three use cases. The reported numbers are for $K=3$. The second column shows how the identification rate falls after the privacy mechanism is applied. The fourth column reports an error metric that is relevant to that use case.} \label{tab:tradeoff_temporal}
\begin{tabular}{|l|c|c|c|c|} \hline
    Mechanism & Identif. Rate & Utility & Impact on Utility & Dataset  \\ \hline
    \begin{tabular}{@{}l@{}}Temporal \\ Downsample\end{tabular} & 3\% $\rightarrow$ 3\% & Gaze Prediction & Avg. Prediction Error Difference = $0.22^\circ$ & DGaze (Not Re-trained)\\ \hline

    \begin{tabular}{@{}l@{}}Temporal \\ Downsample\end{tabular} & 85\% $\rightarrow$ 79\% & AOI Analysis &Dwell Time RMSE = 0.006s &ET-DK2 (360$^\circ$ images)\\ \hline
    \begin{tabular}{@{}l@{}}Temporal \\ Downsample\end{tabular} & 9\% $\rightarrow$ 7\% & Generate Saliency Map & KL-Divergence = 0.0019 & VR-Saliency (360$^\circ$ images) \\ \hline
\end{tabular} 
\end{table*}
\begin{table*}[!h]
\caption{The lowest achievable identification rate\,(IR) for the Spatial Downsample was at $L=64$, and the corresponding impact on utility are reported below. The arrow indicates the IR before and after the privacy mechanism is applied.} \label{tab:tradeoff_spatial}
\begin{tabular}{|l|c|c|c|P{4.0cm}|} \hline
    Mechanism & Identif. Rate & Utility & Impact on Utility & Dataset\\ \hline
    \begin{tabular}{@{}l@{}}Spatial \\ Downsample\end{tabular} & 3\% $\rightarrow$ 2\% & Gaze Prediction & Avg. Prediction Error Difference = $0.51^\circ$ & DGaze\,(Re-trained)\\ \hline
    \begin{tabular}{@{}l@{}}Spatial \\ Downsample\end{tabular} &  85\% $\rightarrow$ 48\% & AOI Analysis & Dwell Time RMSE = 0.2473s & ET-DK2 (360$^\circ$ images)\\ \hline
     \begin{tabular}{@{}l@{}}Spatial \\ Downsample\end{tabular} & 47\% $\rightarrow$ 29\% & Generate Saliency Map& KL-Divergence = 0.1293 & 360\_em (360$^\circ$ videos)\\ \hline
\end{tabular}
\end{table*}

\subsection{Gaze-based Biometric}
\label{sec:biometric_results}
We evaluate the RBF biometric by splitting gaze data from stimuli viewed by each participant into training and testing sets as described in Section\,\ref{sec:evaluation}. For each dataset we evaluate a 75\%/25\%, 50\%/50\%, and 25\%/75\% training/test split, except for DGaze as each participant only saw two stimuli. Identification rate is computed over ten runs with random stimuli selected as part of the training and test set, to account for variance in stimuli content. 

Figure\,\ref{fig:all_results} presents the mean and standard deviation of identification rates for each dataset, along with a baseline rates corresponding to random guessing. For all datasets, identification rate were highest when there was more training data than testing data, i.e., a 75\%/25\% split. ET-DK2 produced the highest identification rate with 85\% on average, where participants viewed 50 static 360$^\circ$ images. VR-Saliency used a similar protocol with 130 participants, however only eight images were shown to each individual on average. A lower identification rate of 9\% was observed in this dataset, compared to a baseline guess rate of 0.77\%. Further analysis comparing identification rates for ET-DK2 using only eight stimuli, and VR-Saliency with eighteen random subjects closed the gap, producing identification rates of 47\% and 22\% respectively. Identification rates for the VR-EyeTracking and 360\_em datasets are lower on average than the ET-DK2 dataset, reporting rates of 33\% and 47\%.
We observed that DGaze produced an identification rate of 2.7\%, showing only slight improvement over a baseline rate of 2.3\%. This dataset differs in that participants moved through two 3D rendered virtual scenes using a controller for teleportation for several minutes at a time, instead of viewing many 360$^\circ$ scenes from a fixed viewpoint. 

In summary, we observe that using more data for training and viewing many different stimuli produces higher identification rates. Thus, it will become easier and easier to re-identify an individual as a large volume of gaze data is collected in a variety of contexts. Identification rates are as high as 85\% depending on the circumstances, highlighting the need to enforce privacy in future mixed reality applications.

Figure\,\ref{fig:privacy_mechanism_results} presents the mean and standard deviations achieved when privacy mechanisms are applied to each dataset. A training/testing split of 75\%/25\% is used to generate these results. We observe that Gaussian noise achieves the most privacy, reducing the identification rate of ET-DK2 from 85\% to 30\% on average. Temporal downsampling is not recommended, as it had the least observed impact on identification rate and event detection is degraded at sampling rates less than 120Hz~\cite{zemblys2018developing}.

\subsection{Utility Evaluation}
\label{sec:utility_results}


The utility of eye-tracking data depends on the context of the application, thus we evaluate the impact of our privacy mechanisms at three different scales: sample-level gaze points, individual-level gaze behavior, and aggregate-level gaze behavior over many individuals. First, we evaluate sample-level utility by computing gaze prediction error using the DGaze neural network architecture, then, individual-level utility by computing dwell time for AOIs defined in the ET-DK2 dataset, and finally, we compute aggregate-level utility measures for generating saliency heatmaps of 360$^\circ$ images and video by computing KL-Divergence for the VR-Saliency, VR-EyeTracking, and 360\_em datasets. Tables \ref{tab:tradeoff_gaussian}, \ref{tab:tradeoff_temporal}, and\,\ref{tab:tradeoff_spatial} present the impact of privacy mechanisms on utility based on the parameter that provided the largest decrease in identification rate.

\noindent \textbf{Gaze Prediction}
Evaluating gaze prediction accuracy involved configuring the DGaze neural network to predict gaze position 100ms into the future, which as a baseline produces an average gaze prediction error of 4.30$^\circ$. 
Gaze prediction error was as high as 9.50$^\circ$ for the Gaussian mechanism, more than double the baseline gaze prediction error reported in~\cite{hu2020dgaze}. Next, we evaluated performance by re-training the DGaze model from scratch and applying privacy mechanisms to both training and testing data dataset. 
This resulted in much lower prediction errors, with results as low as $5.44^\circ$\,(Table\,\ref{tab:tradeoff_gaussian}), which are comparable to the $4.30^\circ$ reported in~\cite{hu2020dgaze}.

Introducing the privacy mechanism to both training and testing data implies that raw gaze data is not shared with any party during model training and deployment. Our experiments indicate that it is still possible to learn a reasonable gaze prediction model without access to the raw gaze data. Withholding raw gaze data from the training dataset is desirable, as it removes the need to safeguard additional data and alleviates the risk of membership inference attacks~\cite{chen2020machine}.
We expect future gaze prediction models will improve in performance, and in turn decrease the absolute gaze prediction error when using gaze data output from the privacy mechanisms.

\noindent \textbf{AOI Analysis}
The impact of privacy mechanisms on area of interest\,(AOI) analysis is measured as the Root Mean Squared Error\,(RMSE) between AOI metrics. There are several popular AOI metrics, suitable for different analyses, such as number of visits to an AOI~\cite{yangandul2018many}, time to first fixation, and number of visits to an AOI~\cite{john2020look}. For an overview of AOI analysis, see the discussion by Le Meur and Baccino~\cite{le2013methods}. For an investigation into privacy mechanisms, we select Dwell Time as a representative AOI metric. Dwell time is the amount of time spent by a user on an AOI, computed as the sum of the durations of all the fixations inside that AOI. The key logical operation is checking whether a fixation location falls within the bounding box that demarcates the AOI, which is the typical first step in all AOI metrics.

If the fixation location is perturbed, such as with the privacy mechanisms proposed above, then we can anticipate an error being introduced in the dwell time computation. We report the RMSE computed between AOI Dwell Time for each individual on the original dataset and after privacy mechanisms are applied, averaged across all stimuli in the dataset.

RMSE in dwell time computation for additive Gaussian noise and temporal downsampling is below 40ms\,(Tables\,\ref{tab:tradeoff_gaussian} and\,\ref{tab:tradeoff_temporal}), which is insignificant for the practical application of AOI metrics, as a fixation itself typically lasts 200ms~\cite{salvucci2000identifying,zemblys2018using}. However, for spatial downsampling, an RMSE of 247ms is introduced, which is greater than the length of one visual fixation. While being a few fixations off on average may not have a large effect on AOI applications such as evidence-based user experience design, it may be noticeable in scenarios with multiple small AOIs close together, such as figuring out which car the user spent longest looking at on a virtual visit to a car dealership.

\noindent \textbf{Saliency Map Generation}
Saliency maps represent a spatial probability distribution of attention over an image or video. Maps are generated by aggregating fixations from eye-tracking data of multiple observers to highlight regions that attract the most attention in the stimulus~\cite{john2019benchmark}. Saliency maps are used directly for gaze prediction~\cite{chen2020salbinet360} and to optimize streaming~\cite{lungaro2018gaze,xu2020state} or rendering~\cite{longhurst2006gpu}. We compute error as the KL-Divergence between a saliency map generated from the original gaze data and the saliency map generated by gaze data after the privacy mechanisms have been applied. KL-Divergence measures the relative entropy between the two saliency maps and is commonly used in loss functions to train deep saliency prediction models and to evaluate learned models~\cite{le2013methods,gutierrez2018introducing,chao2018salgan360,chen2020salbinet360}. 

The spatial errors introduced by the privacy mechanism may cause regions highlighted by the saliency map to shift or spread out, leading to larger KL-Divergence values. A recent survey revealed the best performing model in predicting human fixations produced a KL-Divergence of 0.48 for the MIT300 dataset, with baseline models producing values of 1.24 or higher~\cite{borji2019saliency}. We observed that spatial downsampling produces the largest KL-Divergence on average of 0.1293, while Gaussian and temporal downsampling mechanisms produces much smaller values of 0.0367 and 0.0019 respectively. Spatial downsampling introduced errors that are approximately a fourth of the existing gap in fixation prediction. Errors of this magnitude will cause saliency maps generated from spatially downsampled gaze data to deviate from ground truth, and negatively impact performance of models that use the maps for training. 

%% file: conclusion.tex
\section{Conclusions and Future Work}

As eye-tracking technology is built into mixed reality devices, they open up possibilities for violating user privacy. In this paper, we have examined a specific threat to user privacy: unique user identification based on their eye movement data. This identification would enable colluding applications to connect a user logged in ``anonymously'' with their work ID, for example. 

We first determine biometric identification rates across five datasets of eye movements in immersive environments. We show that identification rates can reach as high as 85\% depending on the type of stimulus used to elicit the eye movements, and the amount of eye movement data collected in total. Our highest identification rates were achieved when viewing many 360$^\circ$ images with short duration\,(ET-DK2), with all datasets having an identification rate higher than chance except DGaze. We hypothesize this is the result of the DGaze dataset providing viewers only two scenes to explore, containing sparse environments with animals that they can follow around by using teleporting to navigate. In the context of saliency Borji~\cite{borji2019saliency} describes the role that stimuli plays in eye movements elicited by viewers, suggesting that datasets from more diverse stimuli is needed to improve generalized performance of saliency prediction models. In the context of privacy, this suggests that the presence of biometric features within gaze data collected in environments differs for photorealistic, static, and dynamic stimuli. Given enough eye movement data collected from the right stimuli, there is an appreciable risk for identification.

We propose a \textit{Gatekeeper} model to alleviate biometric authentication by apps that need AOI metrics or event specific data for their utility. This model provides API calls that return desired metrics and summary information of fixation and saccades to applications without providing streams of raw gaze data, which suffices for certain classes of mixed reality use cases. However, in the case of use cases such as foveated rendering, streaming gaze data is required. We propose that in this case, privacy mechanisms be applied to the raw data stream to reduce identification rate, while maintaining the utility needed for the given application. We evaluated three privacy mechanisms: additive Gaussian noise, temporal downsampling, and spatial downsampling. Our best results used additive Gaussian noise to reduce an identification rate of 85\% to 30\% while supporting AOI analysis, gaze prediction, and saliency map generation.

\noindent \textbf{Implications} Imagine the scenario described earlier of a worker that anonymously attends labor union meetings as User X. The eye-tracking data collected during a VR union meeting attended by User X is exposed through a database breach or collusion with the employer, who then discovers a match between User X and their real identity at a rate greater than chance. Even though they were not the only worker to attend this meeting, biometric data suggested they were the most likely employee to have attended, turning User X into a scapegoat for the entire group. The individual may then have their reputation tarnished in retaliation by their employer. Our investigations are a first step towards protecting such a user. Though the proposed mechanisms lower identification rates, they do not eliminate the possibility of weak identification. More work is needed to create and evaluate mechanisms that allow users, organizations, and platforms to trust eye tracking, and more broadly, behavioral tracking, within mixed reality use cases.

\noindent \textbf{Limitations} Our threat model assumes a trusted platform. In cases where the platform itself cannot be trusted, there is a need for user-implementable solutions, similar in spirit to the user-implementable optical defocus in~\cite{john2020security}. Our characterization of the proposed privacy mechanisms is based on one biometric authentication approach\,(RBFN). As newer methods are developed, we will likely need new privacy mechanisms that can applied as a software patch for the mixed reality headset. This work also considers each privacy mechanism individually. We expect there will be greater gains in terms of privacy when applying a combination of different privacy mechanisms. 

\noindent \textbf{Future Work} In addition to exploring combinations of privacy mechanisms, future work might draw inspiration from research in location privacy, and investigate adapting location k-anonymity schemes for gaze~\cite{gkoulalas2010providing}. It would also be interesting to characterize stimuli as being dangerous from the perspective of biometric signatures, akin to ``click-bait''. More broadly, while our work considers the user privacy, future work might also consider security from a platform's perspective. Consider the case of an attacker injecting gaze positions to fool an AOI metric into thinking that an AOI has been glanced at\,(for monetization of advertisements). One potential solution to this problem is direct anonymous attestation in a trusted platform module\,(TPM) to assure gaze consumers that there have been no injections.